\documentclass[]{elsarticle}

\usepackage{lineno,hyperref}
\usepackage{epsfig}
\usepackage{graphicx}
\usepackage{dcolumn}
\usepackage{amsmath}
\usepackage{bm}
\modulolinenumbers[5]

\journal{}

\begin{document}

\begin{frontmatter}

\title{Next-to-leading order improved perturbative QCD + saturation + hydrodynamics model for $A$+$A$ collisions}

\author[JKL,HIP]{R. Paatelainen}
\author[JKL,HIP]{K.~J. Eskola}
\author[FF]{H. Holopainen}
\author[JKL,HIP]{H. Niemi}
\author[HKI,HIP]{K.~ Tuominen}
\address[JKL]{Department of Physics P.O.Box 35, FI-40014 University of Jyväskyl\"{a}, Finland}
\address[HIP]{Helsinki Institute of Physics, P.O.Box 64, FI-00014 University of Helsinki, Finland}
\address[HKI]{Department of Physics, P.O.Box 64, FI-00014 University of Helsinki, Finland}
\address[FF]{Frankfurt Institute for Advanced Studies, Ruth-Moufang-Str. 1, D-60438 Frankfurt am Main, Germany}

\begin{abstract}

We calculate initial conditions for the hydrodynamical evolution in ultrarelativistic heavy-ion collisions at the LHC and RHIC in an improved next-to-leading order perturbative QCD +  saturation framework. Using viscous  relativistic hydrodynamics, we show that we obtain a good simultaneous description of the centrality dependence of charged particle multiplicities, transverse momentum spectra and elliptic flow at the LHC and at RHIC. In particular, we discuss how the temperature dependence of the shear viscosity is constrained by these data.

\end{abstract}


\end{frontmatter}

\section{Introduction}

In this talk we report the results from the recent studies \cite{Paatelainen:2013at,Paatelainen:2012at}, where we have extended the EKRT model \cite{EKRT} to next-to-leading order (NLO) in perturbative QCD (pQCD), and shown the viability of the model in describing the initial energy density of the produced quark-gluon plasma (QGP) in $A$+$A$ collisions at the LHC and RHIC. Our updated framework combines a rigorous NLO pQCD computation of the minijet transverse energy $(E_T)$ production with the saturation of gluons and hydrodynamics. Latest knowledge of NLO nuclear parton distributions (nPDFs) \cite{EPS09,EPS09s} is utilized. Identifying the key parameters and charting the uncertainties of the model, we obtain a good simultaneous agreement with the charged particle multiplicities and hadron transverse momentum ($p_T$) spectra measured in Au+Au collisions at RHIC and Pb+Pb at the LHC.

The write-up is organized as follows: In Sec.\ 2 we present the updated NLO pQCD + saturation framework for the calculation of the minijet $E_T$. In Sec.\ 3, we discuss the saturation-wise transversally averaged initial conditions in the 5\% most central collisions \cite{Paatelainen:2012at}. Finally, in Sec.\ 4, we discuss the extension to local saturation, which allows us to calculate the initial density profiles at all centralities \cite{Paatelainen:2013at}, and through viscous hydrodynamics, study the temperature dependence of shear viscosity.

\section{Minijet $E_T$ production }

The minijet $E_T$ produced into a rapidity region $\Delta y$ in $A$ + $A$ collisions at an impact parameter $\mathbf{b}$ and above a $p_T$ scale $p_0$, can be computed as 
\begin{equation}
\frac{{\rm d} E_T}{{\rm d}^2 \mathbf{s}} = T_A(\mathbf{s} + \frac{\mathbf{b}}{2})T_A(\mathbf{s} - \frac{\mathbf{b}}{2})\sigma\langle E_T \rangle_{p_0,\Delta y,\beta},
\end{equation}
where $\mathbf{s} = (x,y)$ is the transverse location, and $T_A(\mathbf{s})$ the nuclear thickness function with the Wood-Saxon nuclear density profile. The first $E_T$-moment of the minijet $E_T$ distribution $\sigma\langle E_T \rangle_{p_0,\Delta y,\beta}$ in NLO is computed as \cite{EKL,ET,Paatelainen:2012at}
\begin{equation}
\sigma\langle E_T \rangle_{p_0,\Delta y,\beta} = \sum_{n=2}^{3}\frac{1}{n!}\int [{\rm DPS}]_n \frac{{\rm d}\sigma^{2\rightarrow n}}{[{\rm DPS}]_n}\tilde{S}_{n},
\end{equation}
where the integrations take place in $4-2\epsilon$ spacetime dimensions, and the $2\rightarrow n$ differential partonic cross sections are denoted as ${\rm d}\sigma^{2\rightarrow n}/[{\rm DPS}]_n$. The infrared (IR) and collinear (CL) divergencies present in the partonic NLO cross sections can be regulated by computing the ultraviolet (UV) renormalized squared $2\rightarrow 2$ and $2\rightarrow 3$ scattering matrix elements of order $\alpha_s^3$, in $4-2\epsilon$ dimensions and in the $\overline{\text{MS}}$ scheme \cite{Ellis} (see also \cite{RistoPhD}). In getting from the IR/CL regulated and UV renormalized squared matrix elements to the physical quantities we apply the procedure by S. Ellis, Kunszt and Soper \cite{Kunszt}. The  nPDFs \cite{EPS09,EPS09s} together with the CTEQ6M parton distributions \cite{CTEQ6M:2002} are used in the computation of $\sigma\langle E_T \rangle_{p_0,\Delta y,\beta}$. The measurement functions 
\begin{equation}
\tilde{S}_n = \Theta(\sum_{i=1}^{n}p_{T,i} \geq 2p_0)E_{T,n}\Theta(E_{T,n}\geq\beta p_0),
\end{equation}
where $E_{T,n} = \sum_{i=1}^{n}\Theta(y_i \in \Delta y)p_{T,i}$ and $\Theta$ is the step function, define the total minijet $E_T$ produced in $\Delta y$, and the hard scattering in terms of the minijet transverse momentum $p_{T,i}$ and the cut-off scale $p_0$. The hardness-parameter $\beta$ defines the minimum $E_T$ required in the interval $\Delta y$. As discussed in \cite{Paatelainen:2012at}, any $\beta \in [0,1]$ is acceptable for the rigorous, IR and CL safe, NLO computation.

The saturation criterion for the minijet $E_T$ production in $A$+$A$ collision at non-zero impact parameters is formulated as \cite{Paatelainen:2013at} (see also the discussion in \cite{Paatelainen:2012at,EKT1}) 
\begin{equation}
\label{eq: saturation}
\frac{{\rm d}E_T}{{\rm d}^2\mathbf{s}}(p_0,\sqrt{s_{{\rm NN}}},\Delta y,\mathbf{s},\mathbf{b},\beta) = \frac{K_{\rm sat}}{\pi}p_0^3\Delta y,
\end{equation}
with an unknown proportionality constant $K_{\rm sat} \sim 1$. For given $K_{\rm sat}, \beta$ and cms-energy $\sqrt{s_{NN}}$, we solve the above equation for 
$p_0 = p_{\rm sat}(\sqrt{s_{NN}},A,\mathbf{s},\mathbf{b};K_{\rm sat},\beta)$ and obtain the total ${\rm d}E_T(p_{\rm sat})/{\rm d}^2\mathbf{s}$ produced in a rapidity region $\Delta y$.	

\section{pQCD + average saturation + ideal hydrodynamics}

Let us first discuss the NLO pQCD + averaged (non-local) saturation and ideal hydrodynamics framework introduced in \cite{Paatelainen:2012at}. The average saturation criterion for central $(\mathbf{b}=0)$ $A$+$A$ collisions is obtained by integrating over the transverse plane ${\rm d}^2\mathbf{s}$ in Eq.\ \eqref{eq: saturation}:
\begin{equation}
\label{eq:avesaturation}
E_T(p_0,\sqrt{s_{NN}},\Delta y,\beta) = K_{\rm sat}R_A^2p_0^3\Delta y,
\end{equation}
where $R_A$ is the nuclear radius. Once the average saturation momentum scale $p_0 = p_{\rm sat}(\sqrt{s_{NN}},A;K_{\rm sat},\beta)$ fulfilling the average saturation criterion above is found, the saturated minijet $E_T(p_{\rm sat})$ is converted into the initial QCD matter energy density as
\begin{equation}
\epsilon(\tau_0) = \frac{{\rm d}E_T}{{\rm d}^2\mathbf{s}}\frac{1}{\tau_0\Delta y},
\end{equation}
by assuming that the system thermalizes at formation, $\tau_0=1/p_{\rm sat}$. Furthemore, since the transversally-averaged saturation considered here does not fix the transverse profile for the produced initial energy density, we use either a binary collision (BC) or wounded nucleon (WN) transverse profile. The correlated parameters of the NLO pQCD calculation, $\beta$ and $K_{\rm sat}$, can be fixed based on the measured charged-particle multiplicity at one given cms-energy. When $\beta$ and $K_{\rm sat}$ are fixed, the initial conditions for any other cms-energy can be calculated.

%%%%%%%%%%%%%%%%%%%%% FIGURE %%%%%%%%%%%%%%%%%%%%%%%%%%%%%%%%
\begin{figure}
\center
\epsfxsize 5.5cm \epsfbox{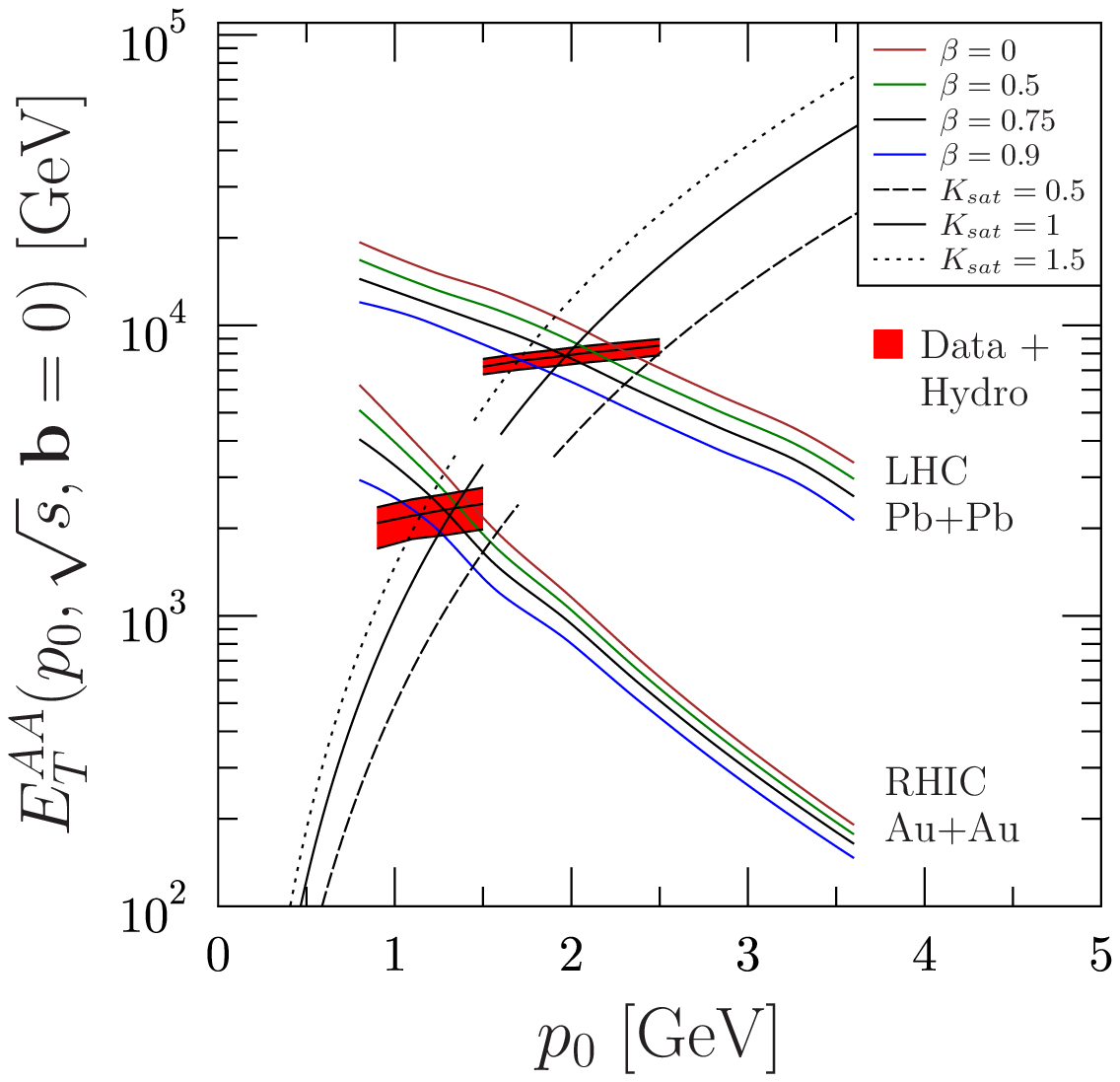} 
\hspace{1.9em}
\epsfxsize 5.5cm \epsfbox{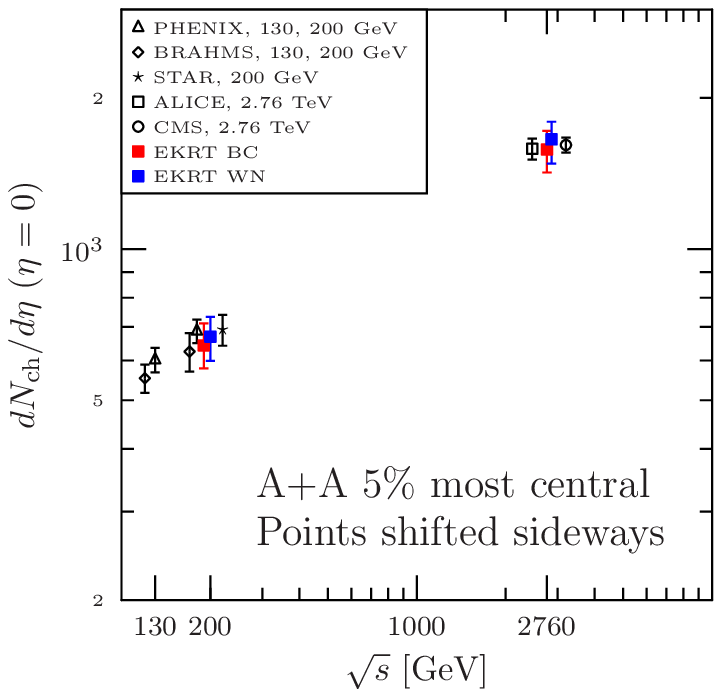} 
\caption{\protect (Color online)
Left panel: The computed NLO minijet $E_T$, as a function of the $p_T$ cut-off $p_0$ (l.h.s.\ of Eq.\ \eqref{eq:avesaturation}). The rising curves are the r.h.s. of Eq.\ \eqref{eq:avesaturation}. Right panel: Computed charged-particle multiplicity ${\rm d}N_{\rm ch}/{\rm d} \eta$ with $\beta = 0.75$ and $K_{\rm sat}=1$, compared with the RHIC and LHC data. From \cite{Paatelainen:2012at}.
}
\label{fig:results1}
\end{figure}
%%%%%%%%%%%%%%%%%%%%% FIGURE %%%%%%%%%%%%%%%%%%%%%%%%%%%%%%%%

In the left panel of Fig.\ \ref{fig:results1} we show the average NLO minijet $E_T$ computed in the mid-rapidity acceptance $\Delta y=1$ with several different $(K_{\rm sat},\beta)$ pairs for the 5 \% most central Au+Au collisions at the RHIC energy $\sqrt{s_{NN}} = 200$ GeV and Pb+Pb collisions at the LHC energy $\sqrt{s_{NN}} = 2.76$ TeV, as a function of the $p_0$ scale. For the implementation of the centrality selection here see \cite{Paatelainen:2012at}. The red bands show the range of values for $E_T$ and $p_0$ that reproduce the measured charged particle multiplicities \cite{Aamodt:2010pb, CMSMUL:2011} (LHC) and  \cite{PHENIXMUL:2005, STARSPECT:2009, BRAHMSMUL:2002} (RHIC) after an ideal-hydrodynamic evolution. Figure \ref{fig:results1} shows directly that there are several different correlated parameter pairs $(K_{\rm sat},\beta)$ that reproduce the measured LHC and RHIC charged-particle multiplicites simultaneously.

\begin{figure}[!]
\epsfxsize 5.9cm \epsfbox{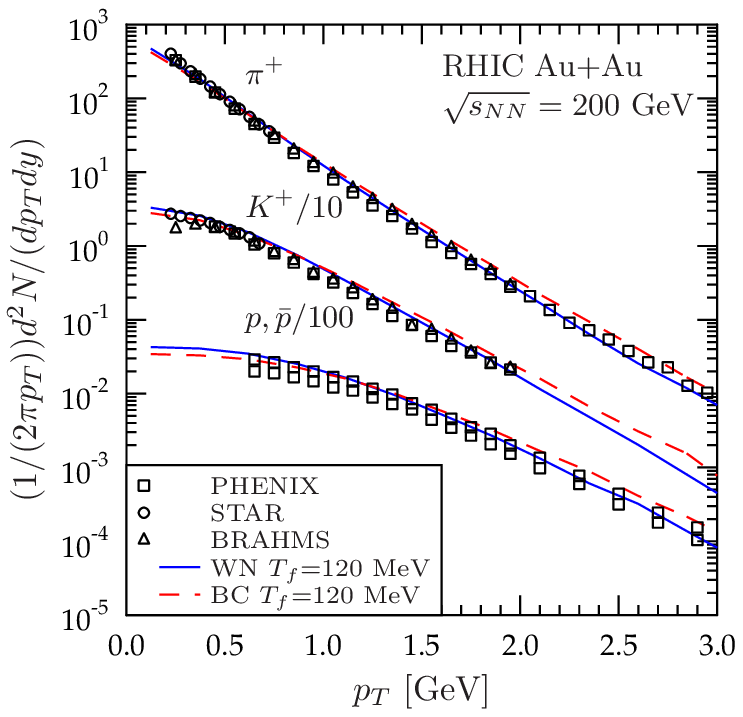} 
\epsfxsize 5.9cm \epsfbox{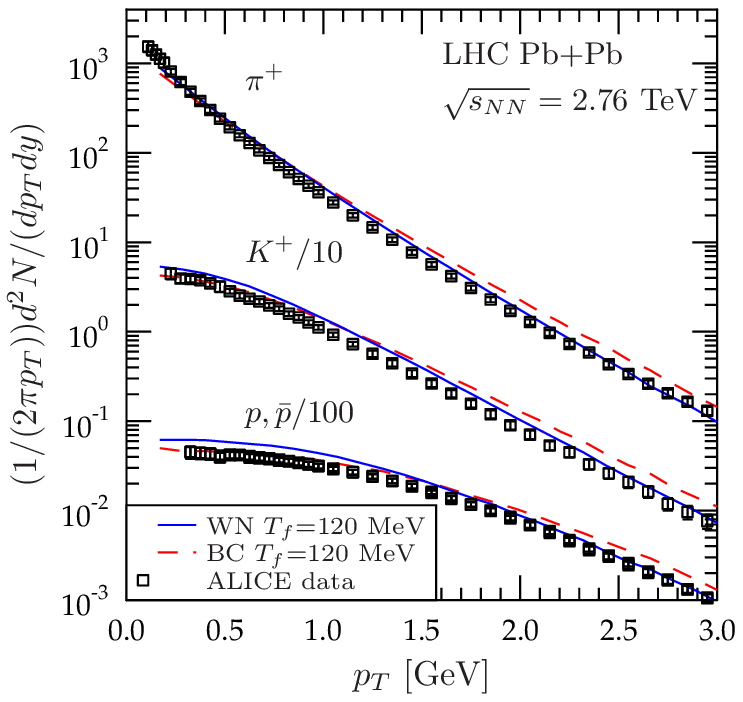} 
\caption{\protect (Color online)
The computed $p_T$ spectra of $\pi^+, K^+, p$ and $\bar p$ at RHIC (Left), compared with the PHENIX, STAR and BRAHMS data, and at the LHC (right), compared with the ALICE data. From \cite{Paatelainen:2012at}.
}
\label{fig:results2}
\end{figure}

Using then one possible parameter combination, $\beta = 0.75$ and $K_{\rm sat}=1$, we show in Fig.\ \ref{fig:results1} (right) the computed charged-particle multiplicity ${\rm d}N_{\rm ch}/{\rm d}\eta$  and in Fig.\ \ref{fig:results2} the $p_T$ spectra of $\pi^+, K^+,p$ and $\bar p$ in 5\% most central Au+Au collisions at RHIC and Pb+Pb collisions at the LHC. Also, the comparison with the data measured  at RHIC \cite{STARSPECT:2009, PHENIXSPECT:2004, BRAHMSSPECT:2004} and at the LHC \cite{ALICESPECT:2012} is shown.

%%%%%%%%%%%%%%%%%%%%% FIGURE %%%%%%%%%%%%%%%%%%%%%%%%%%%%%%%%
%%%%%%%%%%%%%%%%%%%%% FIGURE %%%%%%%%%%%%%%%%%%%%%%%%%%%%%%%%

\section{pQCD + local saturation + viscous hydrodynamics}

Next, we discuss the results obtained in our recent study \cite{Paatelainen:2013at}, where the initial energy density profile for any centrality and impact parameter was computed by using the new pQCD + local saturation + viscous hydrodynamics framework.

Once the solution $p_0 = p_{\rm sat}(\mathbf{s})$ of the transversally local saturation criterion,  Eq.\ \eqref{eq: saturation}, is known for given $K_{\rm sat}$ and $\beta$, the local energy density is obtained as 
\begin{equation}
\epsilon(\mathbf{s},\tau_s) = \frac{{\rm d}E_T}{{\rm d}^2\mathbf{s}\tau_s\Delta y} = \frac{K_{\rm sat}}{\pi}p_{\rm sat}^4,
\end{equation}
where the local formation time is $\tau_s = 1/p_{\rm sat}$. Note that the formation time $\tau_s$ is different at different points in the transverse plane. However, for the hydrodynamical evolution, we need the initial state at a fixed $\tau_0$. For this reason we need to evolve the energy density at all points to the same fixed $\tau_0$. Our strategy is as follows: First, we choose a minimum scale $p^{\rm min}_{\rm sat}=1$ GeV, corresponding a maximum time $\tau_{s	}=1/p^{\rm min}_{\rm sat}$, for which we assume that we can still trust the pQCD calculation. Second, we evolve the energy densities from $\tau_s$ to $\tau_{0}$ using either Bjorken free streaming (FS) or the Bjorken hydrodynamic scaling solution (BJ). We take these two limits to represent the uncertainty in the evolution: In the free streaming case the transverse energy is preserved, while the other limit corresponds the case where a maximum amount of the transverse energy is reduced by the longitudinal pressure. The region below the minimum scale  $p^{\rm min}_{\rm sat}$ is considered as a boundary. To obtain the energy density in this region we use an interpolation, which smoothly connects the FS/BJ-evolved pQCD energy density to the binary profile $\epsilon \propto T_AT_A$ at the dilute edge. For more details  see Ref.\ \cite{Paatelainen:2013at}.

\begin{figure}[!]
\centering
\epsfxsize 5.9cm \epsfbox{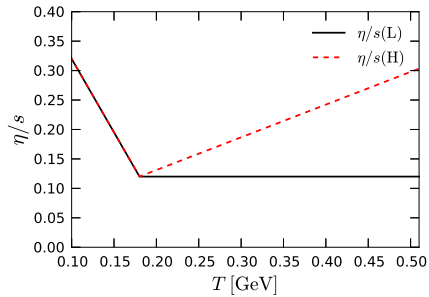} 
\caption{\protect (Color online)
Parametrizations of shear viscosity to entropy density ratio. From \cite{Paatelainen:2013at}.
}
\label{fig:results3}
\end{figure}

For the  hydrodynamical evolution, we take the 2+1 D setup introduced in \cite{Niemi:2011ix}. We use the lattice QCD and hadron resonance gas based equation of state s95p-PCE-v1 \cite{Huovinen:2009yb} with a chemical freeze out temperature $T_{\rm chem} = 175$ MeV. The freeze-out temperature is here always $T_{\rm dec} =100$ MeV. The parametrizations of the temperature dependent shear viscosity $\eta/s(T)$, for which we show the following results, are shown in Fig.\ \ref{fig:results3}. The shear-stress and transverse flow are initially set to zero.

%%%%%%%%%%%%%%%%%%%%% FIGURE %%%%%%%%%%%%%%%%%%%%%%%%%%%%%%%%
%%%%%%%%%%%%%%%%%%%%% FIGURE %%%%%%%%%%%%%%%%%%%%%%%%%%%%%%%%

\begin{figure*}[!]
\epsfxsize 5.9cm \epsfbox{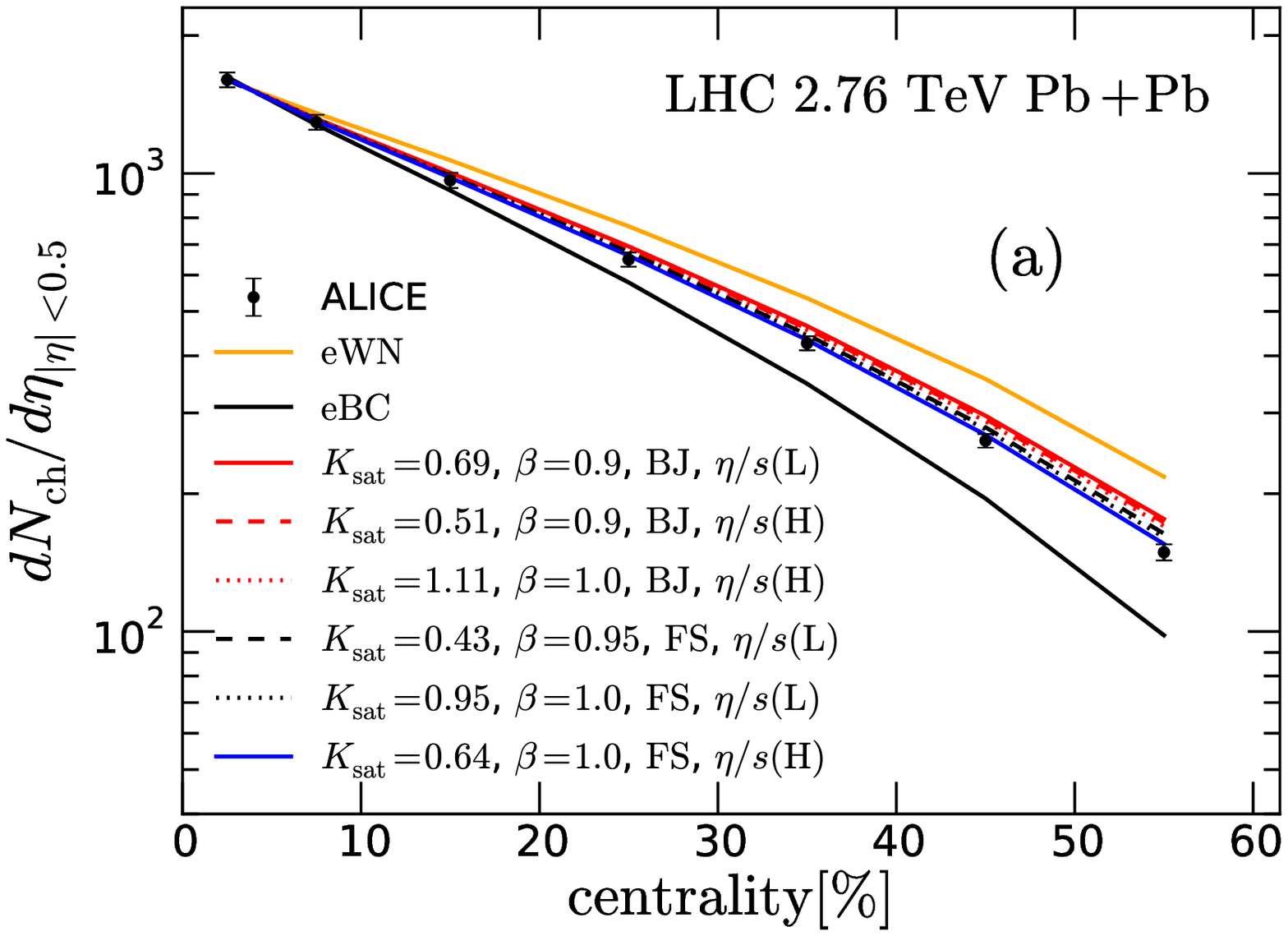} 
\epsfxsize 5.9cm \epsfbox{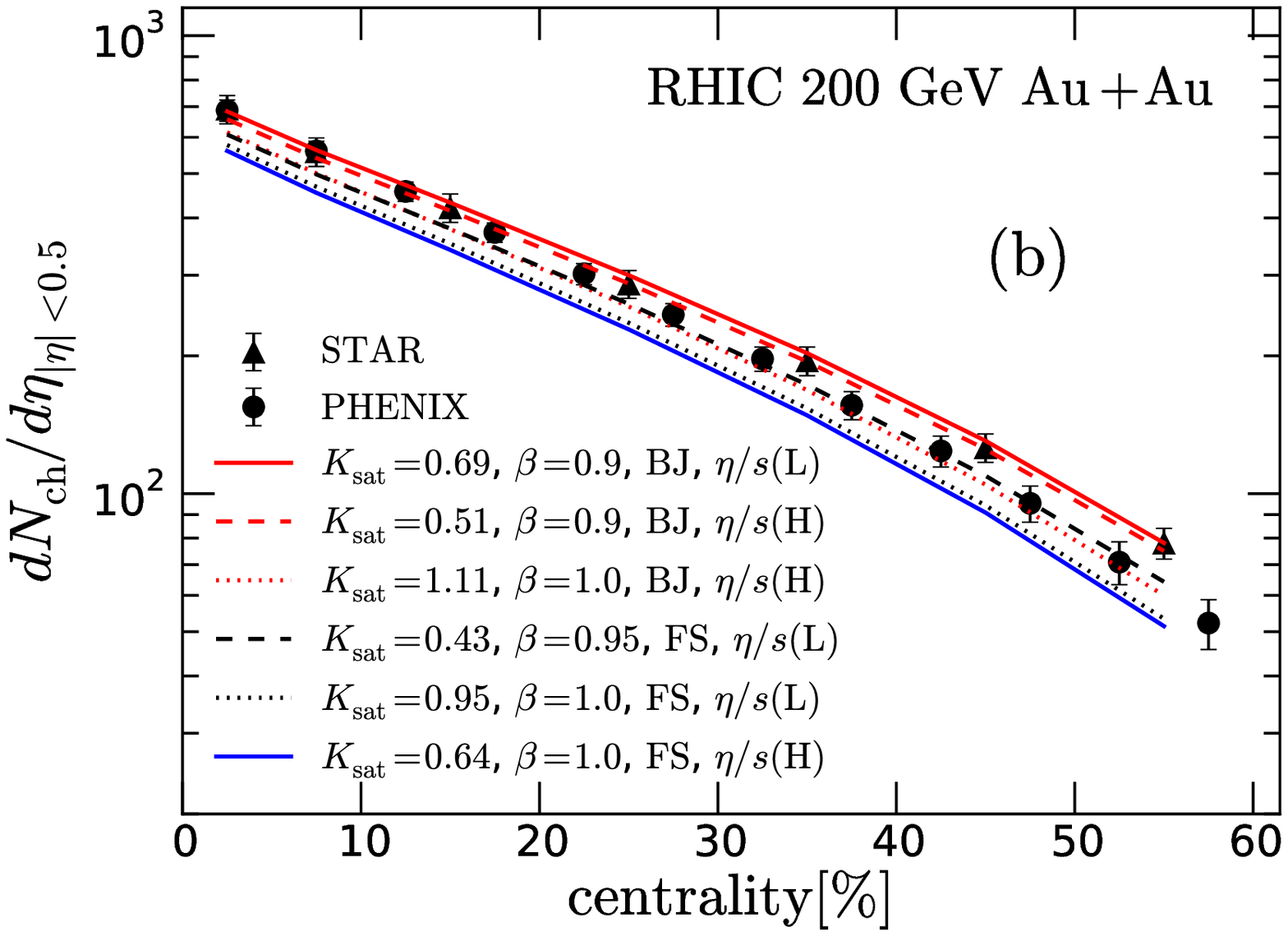} 
\epsfxsize 5.9cm \epsfbox{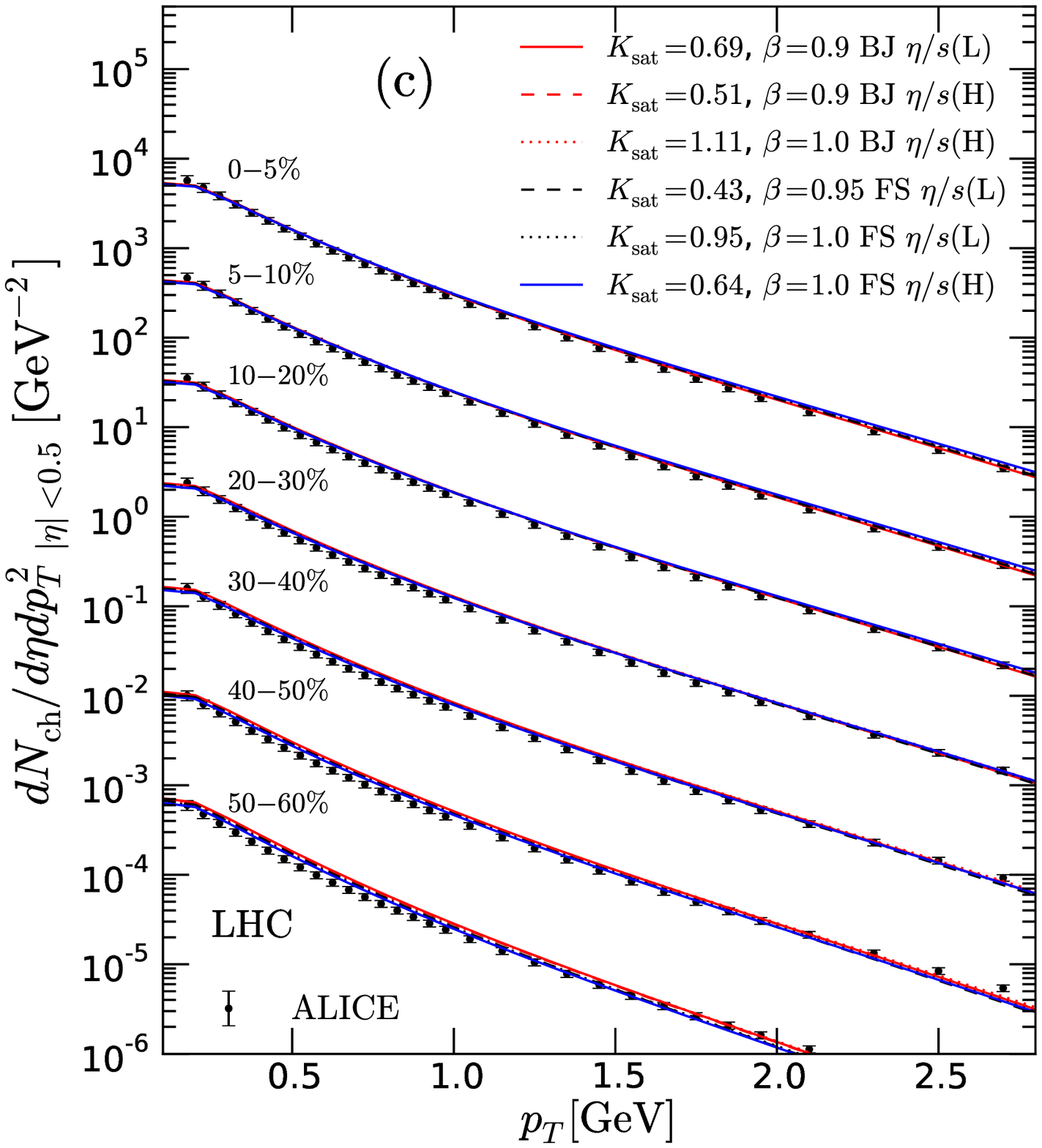} 
\epsfxsize 5.9cm \epsfbox{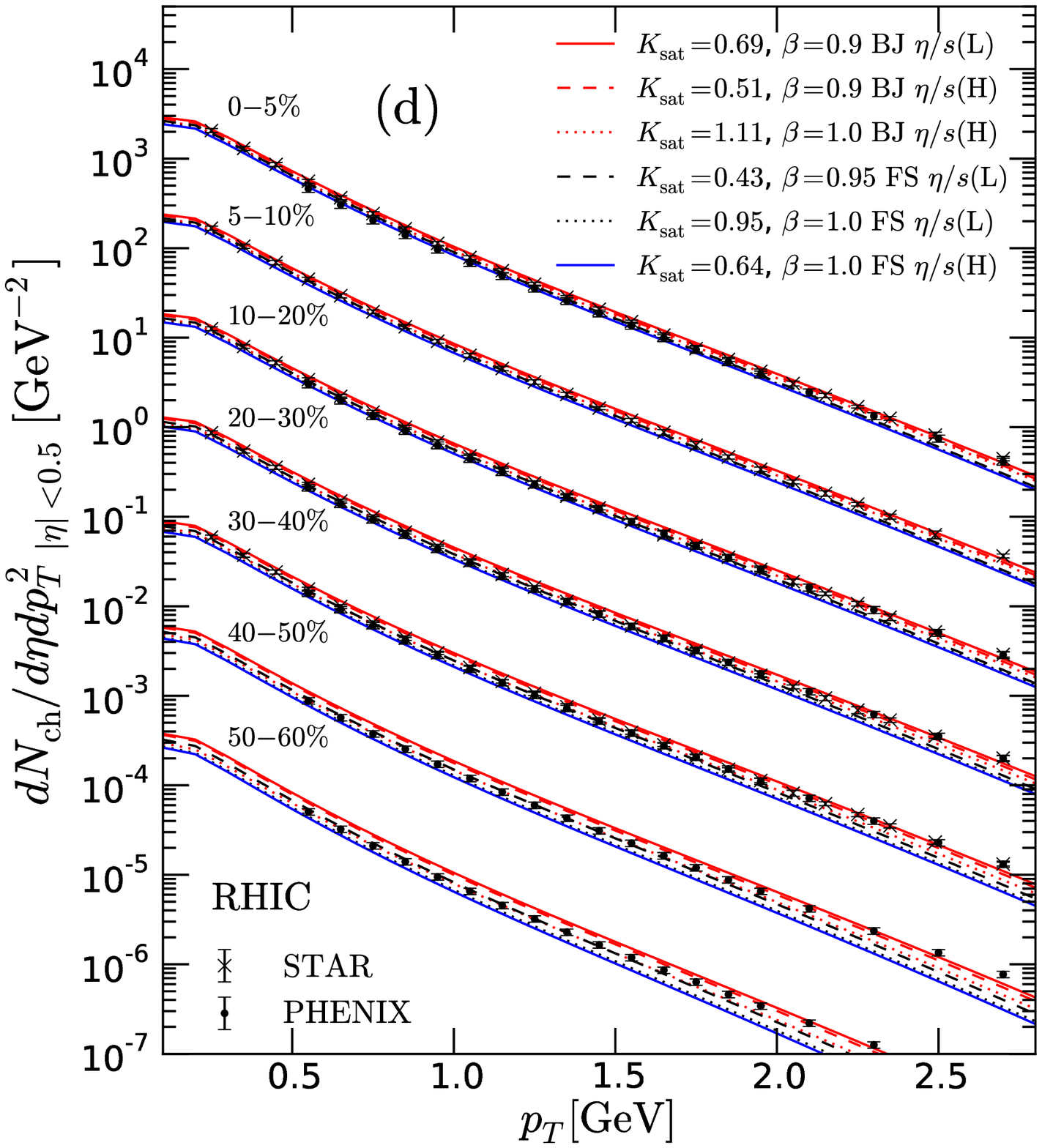} 
\epsfxsize 5.9cm \epsfbox{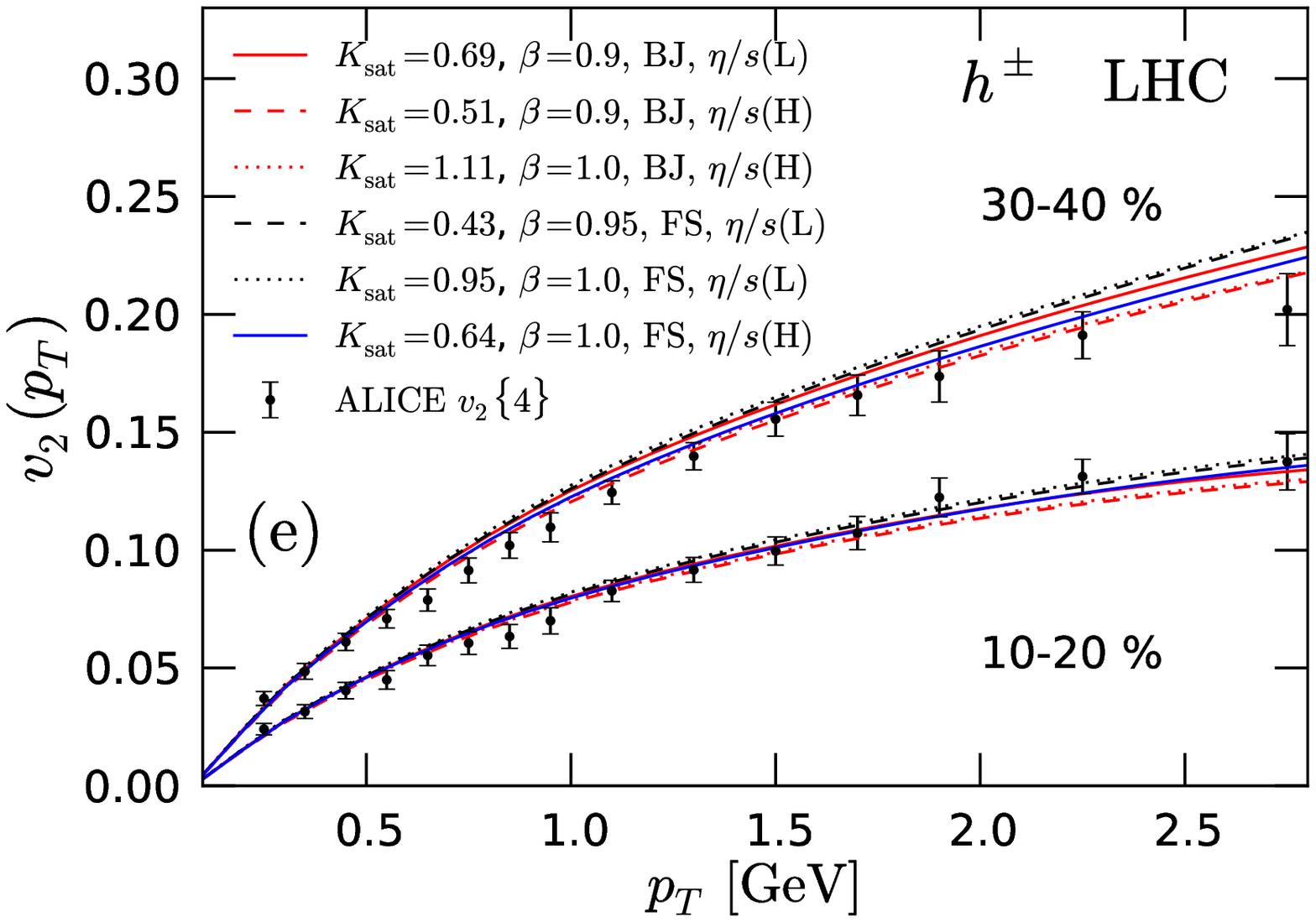} 
\epsfxsize 5.9cm \epsfbox{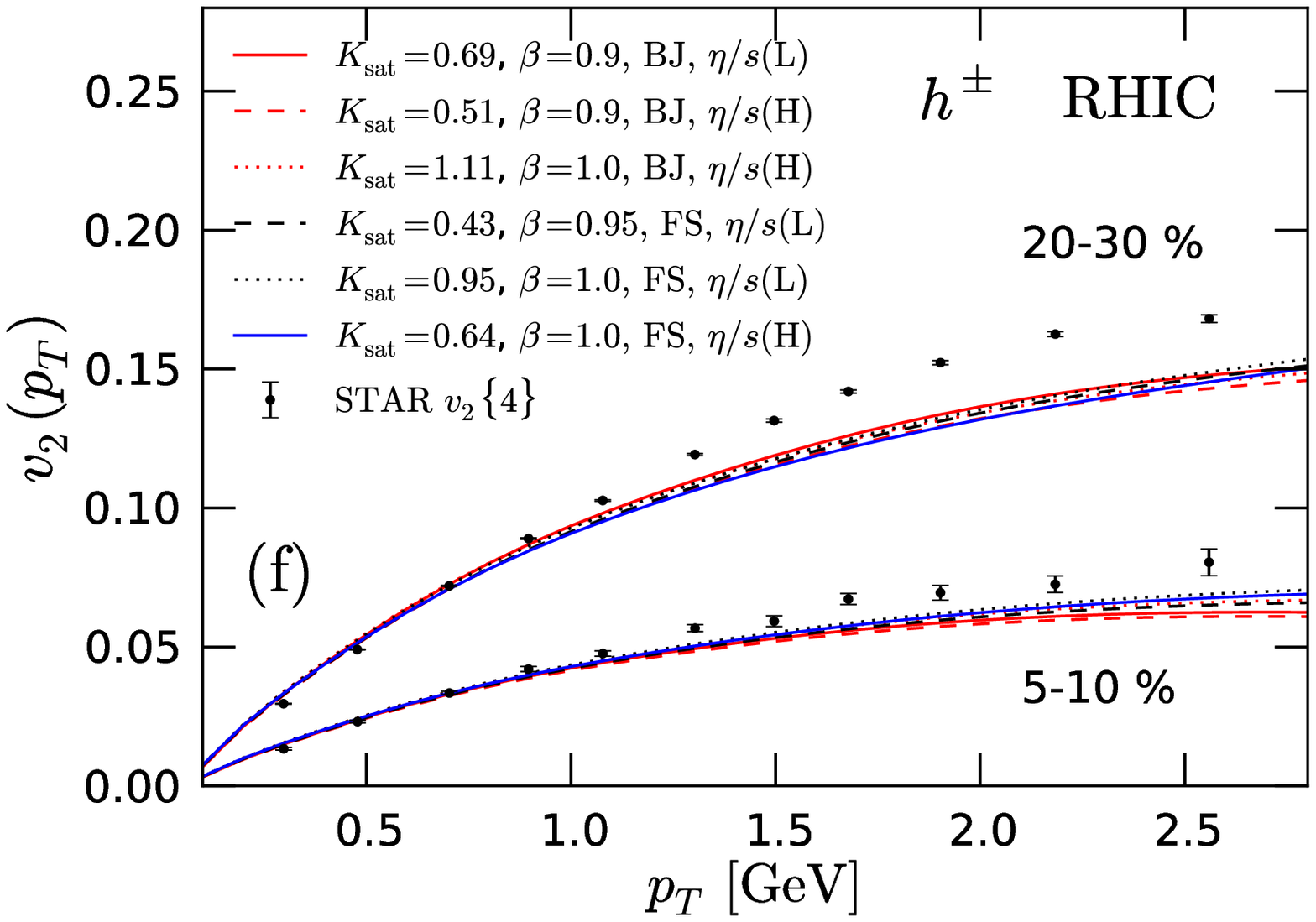} 
\caption{\protect (Color online)
Centrality dependence of the charged hadron multiplicity at the LHC (a) and RHIC (b). 
$p_T$ spectra of charged hadrons at the LHC (c) and RHIC (d), in the same centrality classes as the ALICE data in panel (a), and scaled down by increasing powers of 10.
Elliptic flow coefficients $v_2(p_T)$ at the LHC (e) and RHIC (f), compared with the measured 4-particle cumulant $v_2\{4\}(p_T)$. Labeling of the theory curves in each panel is identical, and the parameter sets $\{K_{\rm sat}, \beta, {\rm BJ/FS}, \eta/s(T) \}$ are indicated. The labels H and L refer to Fig.~\ref{fig:results3}. From \cite{Paatelainen:2013at}.
}
\label{fig: results4}
\end{figure*}

In Fig.\ \ref{fig: results4}a and \ref{fig: results4}b we show the computed centrality dependence of the charged hadron multiplicity in Pb+Pb collisions at $\sqrt{s_{NN}}=2.76$ TeV and in Au+Au collisions at $\sqrt{s_{NN}}=200$ GeV compared with the ALICE (LHC) \cite{Aamodt:2010pb}, PHENIX \cite{PHENIXMUL:2005} and STAR \cite {STARSPECT:2009} (RHIC) data. In practice, the calculation in Figs. \ref{fig: results4}a and \ref{fig: results4}b are performed for each fixed $\{\beta, {\rm BJ}/{\rm FS}, \eta/s(T)\}$, and the remaining parameter $K_{\rm sat}$ is always tuned such that the multiplicity in the 0-5\%  most central collisions at the LHC is reproduced. Next, the obtained centrality dependence of the computed $p_T$-spectra of charged hadrons are shown in Figs. \ref{fig: results4}c for the LHC and in Fig.\ \ref{fig: results4}d for RHIC. The data are from \cite{Abelev:2012hxa} and \cite{Adams:2003kv,Adler:2003au}, correspondingly. Finally, in Figs.\ \ref{fig: results4}e and \ref{fig: results4}f we show the elliptic flow coefficients $v_2(p_T)$ at the LHC and RHIC, respectively. The data are from \cite{Aamodt:2010pa} (LHC) and \cite{Bai} (RHIC).

All the selected parameter combinations give a good description of the $p_T$ spectra simultaneously at the LHC and RHIC. The elliptic flow coefficients depend strongly on the $\eta/s(T)$ parametrization: an ideal fluid description would not give the correct $v_2(p_T)$, while with both $\eta/s(T)$ parametrizations considered here we get a good agreement with the data. Before engaging in a more complete global analysis for $\eta/s(T)$, the initial event-by-event density fluctuations need to be considered in this framework. This is work in progress.

%%%%%%%%%%%%%%%%%%%%% FIGURE %%%%%%%%%%%%%%%%%%%%%%%%%%%%%%%%
%%%%%%%%%%%%%%%%%%%%% FIGURE %%%%%%%%%%%%%%%%%%%%%%%%%%%%%%%%

\section*{Acknowledgments}

This work was financially supported by the Jenny and Antti Wihuri Foundation (RP) and the Academy projects of Finland 133005 (KJE) and 267842 (KT).  We thank CSC-IT Center Science for supercomputing time.

%\section*{References}


\begin{thebibliography}{99}


\bibitem{Paatelainen:2013at}
	R.~Paatelainen, K.~J.~Eskola, H.~Niemi and K.~Tuominen,
	arXiv:1310.3105 [hep-ph], submitted to Phys.\ Lett.\ B.


\bibitem{Paatelainen:2012at}
	R.~Paatelainen, K.~J.~Eskola, H.~Holopainen and K.~Tuominen,
  Phys.\ Rev.\ C {\bf 87}, 044904 (2013).
  

\bibitem{EKRT} 
  K.~J.~Eskola, K.~Kajantie, P.~V.~Ruuskanen and K.~Tuominen,
  Nucl.\ Phys.\ B {\bf 570}, 379 (2000).


\bibitem{EPS09} 
  K.~J.~Eskola, H.~Paukkunen and C.~A.~Salgado,
  JHEP {\bf 0904}, 065 (2009).


\bibitem{EPS09s}
I.~Helenius, K.~J.~Eskola, H.~Honkanen and C.~A.~Salgado,
\newblock JHEP {\bf 1207}, 073 (2012).



\bibitem{EKL}
 K.~J.~Eskola, K.~Kajantie and J.~Lindfors,
 Nucl.\ Phys.\ B {\bf 323}, 37 (1989).

\bibitem{ET}
 K.~J.~Eskola and  K.~Tuominen,
 Phys.\ Lett.\ B {\bf 489}, 329 (2000); Phys.\ Rev.\ D {\bf 63}, 114006 (2001).  
 
\bibitem{Ellis}
R.~K.~Ellis and J.~C.~Sexton,
\newblock Nucl. Phys. B {\bf 269}, 445 (1986).


\bibitem{RistoPhD}
R.~Paatelainen, PhD thesis, in progress.


\bibitem{Kunszt}
Z.~Kunszt and D.~E.~Soper,
\newblock Phys. Rev. D {\bf 46}, 192 (1992).    


\bibitem{CTEQ6M:2002}
 J.~Pumplin {\it et al.}
 JHEP  {\bf 0207}, 012 (2002).

\bibitem{EKT1} 
  K.~J.~Eskola, K.~Kajantie and K.~Tuominen,
  Phys.\ Lett.\ B {\bf 497}, 39 (2001).


\bibitem{Aamodt:2010pb} 
  B.~Abelev {\it et al.}  [ALICE Collaboration],
  Phys.\ Rev.\ Lett.\  {\bf 105}, 252301 (2010).

\bibitem{CMSMUL:2011}
 S.~Chatrchyan {\it et al.}  [CMS Collaboration],
 JHEP {\bf 1108}, 141 (2011).

\bibitem{PHENIXMUL:2005}
 S.~S.~Adler {\it et al.} [PHENIX Collaboration] ,
 Phys.\ Rev. C {\bf 71}, 034908 (2005).


\bibitem{STARSPECT:2009} 
 B.~Abelev {\it et al.} [STAR Collaboration],
 Phys.\ Rev. C {\bf 79}, 034909 (2009).

\bibitem{BRAHMSMUL:2002}
 I.~G.~Bearden {\it et al.} [BRAHMS Collaboration], 
 Phys.\ Rev. Lett. {\bf 88}, 202301 (2002).


\bibitem{PHENIXSPECT:2004}
 S.~S.~Adler {\it et al.} [PHENIX Collaboration],
 Phys.\ Rev. C {\bf 69}, 034909 (2004).


\bibitem{BRAHMSSPECT:2004}
I.~G.~Bearden {\it et al.} [BRAHMS Collaboration],  
Phys.\ Rev. Lett. {\bf 94}, 162301 (2005).



\bibitem{ALICESPECT:2012} 
  B.~Abelev {\it et al.}  [ALICE Collaboration],
  %``Pion, Kaon, and Proton Production in Central Pb--Pb Collisions at $\sqrt{s_{NN}} = 2.76$ TeV,''
  Phys.\ Rev.\ Lett.\  {\bf 109}, 252301 (2012).
  %%CITATION = ARXIV:1208.1974;%%
  %40 citations counted in INSPIRE as of 15 Jan 2014


\bibitem{Niemi:2011ix} 
  H.~Niemi, G.~S.~Denicol, P.~Huovinen, E.~Molnar and D.~H.~Rischke,
  %``Influence of the shear viscosity of the quark-gluon plasma on elliptic flow in ultrarelativistic heavy-ion collisions,''
  Phys.\ Rev.\ Lett.\  {\bf 106}, 212302 (2011).



\bibitem{Huovinen:2009yb} 
  P.~Huovinen and P.~Petreczky,
  Nucl.\ Phys.\ A {\bf 837}, 26 (2010).


\bibitem{Abelev:2012hxa} 
  B.~Abelev {\it et al.}  [ALICE Collaboration],
  %``Centrality Dependence of Charged Particle Production at Large Transverse Momentum in Pb--Pb 
  % Collisions at $\sqrt{s_{\rm{NN}}} = 2.76$ TeV,''
  Phys.\ Lett.\ B {\bf 720}, 52 (2013).
	%  [arXiv:1208.2711 [hep-ex]]. 
	%%CITATION = ARXIV:1208.2711;%% 


\bibitem{Adams:2003kv} 
  J.~Adams {\it et al.}  [STAR Collaboration],
  %``Transverse momentum and collision energy dependence of high p(T) hadron suppression in Au+Au 
  % collisions at ultrarelativistic energies,''
  Phys.\ Rev.\ Lett.\  {\bf 91}, 172302 (2003).
	%  [nucl-ex/0305015]. 
	%%CITATION = NUCL-EX/0305015;%% 
    
\bibitem{Adler:2003au} 
  S.~S.~Adler {\it et al.}  [PHENIX Collaboration],
  %``High $p_{T}$ charged hadron suppression in Au + Au collisions at $\sqrt{s}_{NN} = 200$ GeV,''
  Phys.\ Rev.\ C {\bf 69}, 034910 (2004).
	%  [nucl-ex/0308006].
	%%CITATION = NUCL-EX/0308006;%%

\bibitem{Aamodt:2010pa} 
  K.~Aamodt {\it et al.}  [ALICE Collaboration],
  %``Elliptic flow of charged particles in Pb-Pb collisions at 2.76 TeV,''
  Phys.\ Rev.\ Lett.\  {\bf 105}, 252302 (2010).
	%  [arXiv:1011.3914 [nucl-ex]].
  %%CITATION = ARXIV:1011.3914;%%
  
\bibitem{Bai}
 	Y.~Bai, Ph.D. Thesis, Nikhef and Utrecht University, The Netherlands (2007);
	% \bibitem{Tang:2008if}
  A.~Tang  [STAR Collaboration],
  %``Flow Results and Hints of Incomplete Thermalization,''
  arXiv:0808.2144 [nucl-ex].
  %%CITATION = ARXIV:0808.2144;%%


%\bibitem{Adler:2004zn} 
 % S.~S.~Adler {\it et al.}  [PHENIX Collaboration],
  %``Systematic studies of the centrality and s(NN)**(1/2) dependence of the 
  % d E(T) / d eta and d (N(ch) / d eta in heavy ion collisions at mid-rapidity,''
 % Phys.\ Rev.\ C {\bf 71}, 034908 (2005)
 % [Erratum-ibid.\ C {\bf 71}, 049901 (2005)].
	%  [nucl-ex/0409015].
	%%CITATION = NUCL-EX/0409015;%%


  
 
 
 
\end{thebibliography}
\end{document}